\documentclass[%
 reprint,
 amsmath,amssymb,
 aps,
]{revtex4-2}
\usepackage{xcolor}
\usepackage{graphicx}
\usepackage{dcolumn}
\usepackage{bm}

\begin{document}

\title{A Minimal Nonlocal Theory of Thixotropic Flow}

\author{Saghar Zolfaghari}
\affiliation{Department of Mechanical and Industrial Engineering, Northeastern University, Boston, MA, USA.}

\author{Safa Jamali}
\email{s.jamali@northeastern.edu}
\affiliation{Department of Mechanical and Industrial Engineering, Northeastern University, Boston, MA, USA.}

\date{\today}

\begin{abstract}

Dense amorphous materials exhibit both nonlocal flow cooperativity and pronounced history dependence, yet existing continuum models capture only one of these features at a time. Nonlocal rheologies are intrinsically memoryless, while thixotropic models remain local. Here we introduce a coupling between structural memory and nonlocal fluidity to include aging and rejuvenation in nonlocal granular fluidity. The resulting model reproduces hysteresis in shear-rate sweeps and delayed yielding in creep, while preserving nonlocal flow profiles. By introducing memory-augmented non-local granular fluidity (MNGF), we show that non-locality alone cannot encode history, and memory alone cannot encode spatial cooperativity, but their coupling is essential—and minimal. These results demonstrate that memory and nonlocality must be treated jointly to describe history-dependent flows, and provide a unified framework for modeling time-dependent rheology in dense amorphous materials.

\end{abstract}

\maketitle



Yield stress fluids and granular materials are among the most common substances on Earth, and also essential in a wide range of industries such as construction, energy, food, and pharmaceuticals~\cite{Henann2013, Berzi2024, Bonn2017YieldStress, BonnDenn2009Science, JoshiPetekidis2018YieldStress, ForterrePouliquen2008ARFM, JerolmackDaniels2019SoftMatterEarth}. Their behavior ranges from solid-like resistance at rest to fluid-like flow under stress, showing rich and diverse phenomenology. To unify the solid-like and fluid-like descriptions, a major advance was the development of the inertial rheology, or $\mu(I)$ framework, which relates the stress ratio $\mu = \sigma/P$ to the inertial number $I = \dot{\gamma} d \sqrt{\rho_s/P}$, where $d$ is the grain size, $\rho_s$ the solid density, and $P$ the confining pressure~\cite{Bagnold1954, GDRMiDi2004}. This description captures dense granular flows under uniform shear and rapid-flow conditions, but breaks down in slow, quasi-static, or heterogeneous regimes, where experiments reveal flow below the apparent yield, wide shear bands, exponential velocity decay into static zones, and falls short of recovering sub-yield creep, finite shear-band widths, or geometry-dependent flow fields~\cite{daCruz2005, Jop2006,Kamrin2015Nonlocal, Kamrin2014, Mohan2002, Hashiguchi2007, Pouliquen2009, Savage1998, Jenkins2010, Aranson2002, Bouzid2013, Kamrin2007, Henann2014IJPlas}. Nonlocal Granular Fluidity (NGF) model, which introduces a scalar fluidity field as a measure of the material’s capacity to flow under stress~\cite{kamrin2019} was developed as an extension of $\mu(I)$ rheology, to address these limitations. In NGF, fluidity evolves according to a Laplacian diffusion term scaled by a cooperativity length proportional to the grain size, allowing the model to reproduce rheological features including flow in sub-yield regions, finite shear-band thickness, size-dependent strengthening, and secondary rheology, whereby shear in one region induces creep in an otherwise static zone~\cite{Kamrin2012, Kamrin2015Nonlocal, Henann2014, Kamrin2024, Henann2014IJPlas}.

In addition to non-locality, many amorphous particulate materials exhibit pronounced time-dependent rheology characterized by a decrease of viscosity under shear and recovery at rest~\cite{Mewis2009, Jamali2022, Larson2015Thixotropy, Bonn2017YieldStress,Coussot2002Avalanche, Coussot2002Viscosity, Benzi2021, Moorcroft2011, Divoux2011,Divoux2013, Radhakrishnan2017, Jamali2019,Jaeger1996, Bocquet2009}. By contrast, the NGF represents an essentially ideal viscoplastic fluid~\cite{Kamrin2012, Henann2013, ForterrePouliquen2008ARFM, Saramito2016, Balmforth2014Yielding, Martin2017GranularCollapse}. Non-local models generally lack a physical connection to the lively nature of heterogeneous contact networks ~\cite{Nabizadeh2024NetworkPhysics, Bantawa2023HiddenHierarchy, ColomboDelGado2014StressLocalization, Tang2018AnnularNonlocal, Berthier2019FailureForecasting, Amon2017ImagingGranularPreface}. The continuous breakage and re-formation of contact chains and networks under flow lead to highly heterogeneous and intermittent local rearrangements that are also observed in other dense particulate systems~\cite{Coussot2014, Bocquet2009, Bassett2015ForceChainNetworks, Peters2005ForceChains, Azema2012ForceChains, Mangal2024SmallVariations, Nabizadeh2022ForceClusters}. When left at rest, the particulate network structures reorganize and strengthen, leading to an increase in elasticity and resistance to flow. Under applied shear, however, internal structures breaks down, resulting in overall increased fluidity~\cite{Coussot2002Viscosity, Divoux2011, Fielding2014}. Rheologically, this is characterized as thixotropy, manifesting in stress overshoots during shear startup~\cite{Moorcroft2011}, hysteresis ~\cite{Jamali2019, Divoux2013, Radhakrishnan2017}, viscosity bifurcation and delayed yielding after flow cessation~\cite{Bocquet2009, Benzi2021, Benzi2023}. Nonetheless, most thixotropic constitutive models remain local in their description ~\cite{Jamali2022, Larson2019, Coussot2002Avalanche, Bonn2017YieldStress}.

In this work, we propose and analyze a minimal memory–nonlocal coupling: an augmented nonlocal fluidity model that incorporates an internal structural parameter to represent aging and rejuvenation. Note that the goal here is not to introduce a new rheology, but to minimally extend NGF so that the flow history can influence nonlocal fluidization. The granular fluidity, denoted by $g$, in NGF is defined as the ratio of shear rate to shear stress, and evolves according to a diffusion-like partial differential equation~\cite{Kamrin2012, Henann2013}:

\begin{equation}
    t_0 \,\dot{g} = A^2 d^2 \nabla^2 g - \Big[(\mu_s - \mu)g + b \sqrt{\frac{\rho_s d^2}{P}} \,\mu g^2\Big],
    \label{eq:ngf}
\end{equation}

where $t_0$ is a characteristic time scale, $A$ is a dimensionless nonlocal amplitude, $d$ is the particle diameter, $b$ is a dimensionless material parameter controlling rate dependence, $\mu$ is the stress ratio, $\mu_s$ is the static yield threshold, $\rho_s$ is the particle density, and $P$ is the pressure. In eq.\ref{eq:ngf}, the fluidity field $g$ responds instantaneously to changes in the stress ratio $\mu$. As a result, the model lacks any kinematic memory of the flow history. In thixotropic models, \textit{structural parameter}, typically denoted $\lambda$, a coarse-grained scalar encoding the local structural state is generally sufficient to capture aging and rejuvenation without resolving microscopic details~\cite{Mewis2009, Larson2019}. The kinetics of $\lambda$ are usually written with two competing terms: one describing \textit{build-up}, i.e.\ the spontaneous formation of structure, and another describing \textit{break-up}, the shear-induced destruction of structure ~\cite{Coussot2002Viscosity, Jamali2022, Larson2019, Beris2021BloodRheology}.

To augment flow memory into the non-local fluidity model, we introduce a single additional coupling term, allowing the structural state to modulate fluidity. This is arguably the simplest modification required to produce history dependence. The coupled time evolution equations of $\lambda$ and $g$ together yield a \textit{Memory-augmented Non-local Granular Fluidity} (MNGF) model:

\begin{align}
    \dot{\lambda} &= \frac{1}{\tau}(1 - \lambda) - \alpha \lambda \dot{\gamma},
    \qquad
    \dot{\gamma} = \mu g, \label{eq:lambda_final} \\
    t_0 \dot{g} &= A^2 d^2 \nabla^2 g - \big[(\mu_s - \mu)g
+ b \sqrt{\frac{\rho_s d^2}{P}}\,\mu g^2\big]
    \nonumber \\
    &\quad + a_{\text{age}}\left(g e^{-\beta \lambda} - g\right),
    \label{eq:g_final}
\end{align}

Equation~\eqref{eq:lambda_final} describes the evolution of the structural parameter $\lambda$, with $\mu g$ providing the local shear rate through the constitutive relation, thereby ensuring that stronger flows accelerate rejuvenation. Physically, $\lambda$ quantifies the extent of contact network formation within the granular assembly. A large value of $\lambda$ corresponds to a well-developed network of force chains and contacts, associated with higher resistance to flow, while a small value of $\lambda$, corresponds to a fluidized structure that flows more easily. In eq.~\eqref{eq:lambda_final}  the first and second terms represent formation and destruction of the structure, respectively. The the rate of flow-driven destruction is naturally proportional to the macroscopic deformation rate, $\dot{\gamma}$. $\tau$ is the characteristic timescale for the particulate contact network to be fully restored. This form implies that $\lambda$ relaxes toward unity at rest, and decays to zero under large deformations. Equation~\eqref{eq:g_final} governs the dynamics of the granular fluidity $g$. The first three terms on the right-hand side correspond to the standard NGF model. The final term, proportional to $a_{\text{age}}(g e^{-\beta \lambda}-g)$, couples the fluidity field to the structural parameter $\lambda$. This term captures how the local structural state modifies the effective fluidity. $\beta$ controls how sensitive the fluidity is to $\lambda$. In small $\beta$ regime, fluidity persists and the material remains relatively soft, consistent with weakly aging systems. For moderate values of $\beta=0.1$–$1$, flow persists but is increasingly reduced as $\lambda$ grows. For high values of $\beta>10$, the suppression of fluidity is extremely strong, and $g$ approaches zero even when the structure is not fully developed.

\begin{figure}[htbp]
    \centering
    \includegraphics[width=\columnwidth]{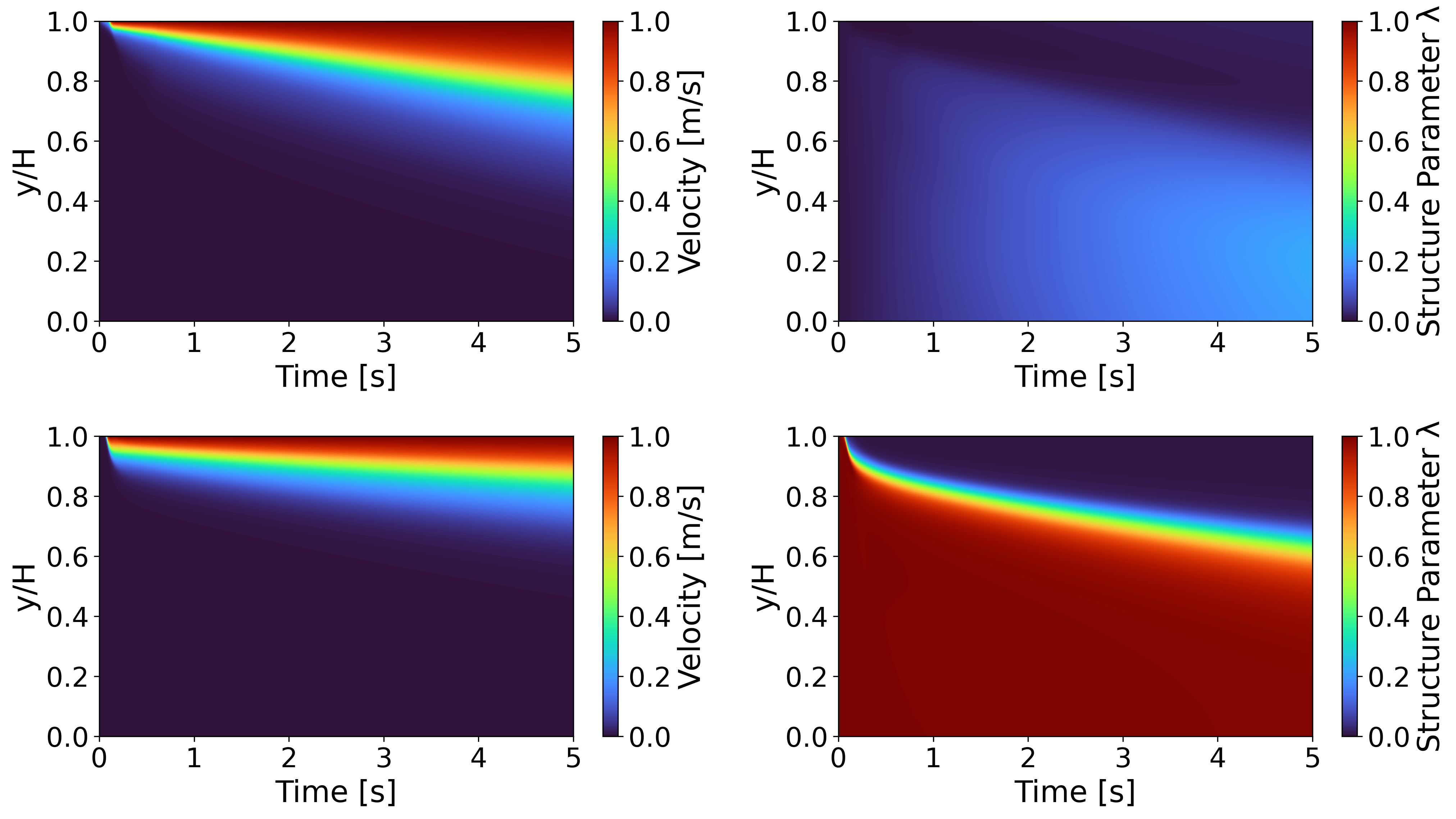}
    \caption{Kymographs of the flow velocity (left column) and structure parameter $\lambda$ (right column) during planar shear flow. Top row shows an initially fully destructured state ($\lambda \approx 0$), and the bottom row shows a fully structured initial condition ($\lambda \approx 1$). Simulations are performed with parameters $t_0 = 0.01\,\text{s}$, $A = 0.84$, $d = 8.0\times 10^{-4}\,\text{m}$, $\mu_s = 0.29$, $\rho = 2450\,\text{kg}/\text{m}^3$, $P = 10\,\text{kPa}$, $\beta = 0.1$, $a_{\text{age}} = 5.0 \,\text{s}^{-1}$, $\tau = 20.0\,\text{s}$, $\alpha = 1.0$, and $b = 1.5$.}
    \label{fig:velocity-structure-profiles}
\end{figure}

Additionally, $a_{\text{age}}$ sets the timescale over which the structural suppression acts. When $a_{\text{age}}$ is small ($0.1$–$1.0$), aging is slow and fluidity persists, which produces weak memory effects. For intermediate values ($1.0$–$5.0$), aging and rejuvenation are balanced: fluidity decays at rest but recovers under shear, leading to time-dependent hysteresis loops that are stable and reproducible. For very large values ($a_{\text{age}}>5.0$), aging dominates and fluidity drops rapidly, producing almost immediate solidification. In this limit, the material can become effectively arrested, consistent with the rapid aging behavior observed in dry granular systems. In MNGF formalism, the stress is both history-dependent, through the evolution of $\lambda$, and inherently nonlocal, through the diffusion of $g$, reflecting cooperative rearrangements and structural aging. These governing constitutive equations are easily coupled with continuity and momentum conservation, and solved for a planar shear flow confined between two parallel plates separated by a gap of height $H$, in which the bottom plate is stationary and the top plate is driven at a constant velocity, with results shown in Fig.~\ref{fig:velocity-structure-profiles}.

Firstly, results in Fig.~\ref{fig:velocity-structure-profiles} show distinct dynamics for velocity and structure depending on the initial condition. In the fully destructured case (top row), the system begins in a soft state with low structural parameter values. The velocity rapidly penetrates into the bulk, producing a flowing layer that thickens with time. Simultaneously, the structural parameter $\lambda$ gradually rebuilds from low values, especially near the lower wall where the deformation rate is close to zero. In contrast, the fully structured case (bottom row) begins in a rigid state with $\lambda \approx 1$ throughout the domain. The initial velocity penetration is therefore much more localized near the moving wall, and the bulk remains essentially solid-like. Over time, shear-induced rejuvenation reduces $\lambda$ in the upper region, allowing the velocity to propagate deeper into the material.

\begin{figure*}[htbp]
    \centering

    \includegraphics[
        width=0.95\textwidth,
        trim=10 15 0 15,
        clip
    ]{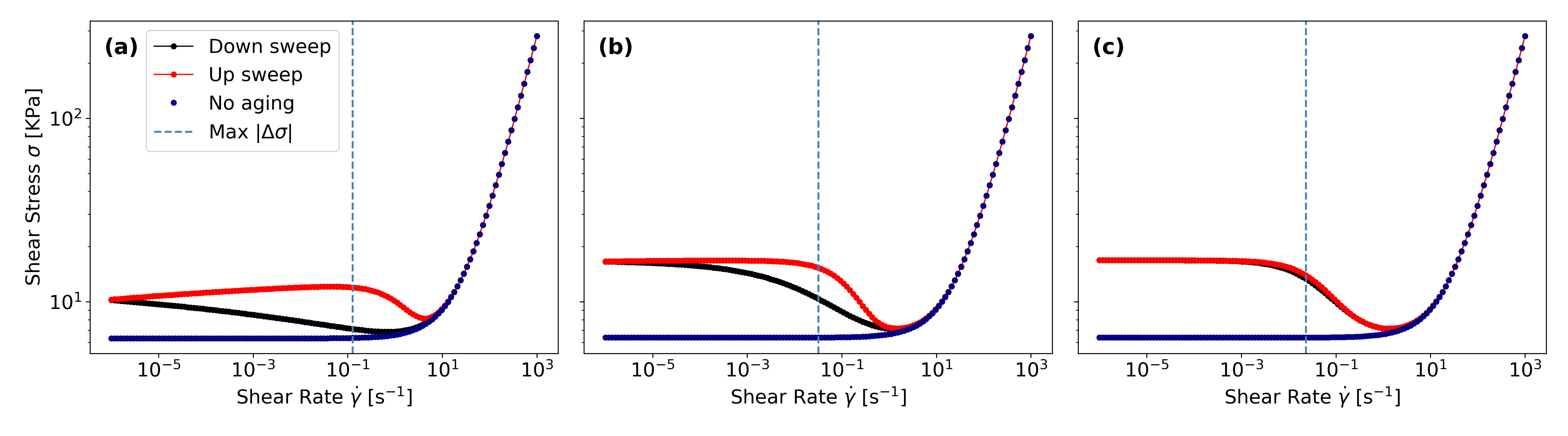}

    \vspace{1em}

    \includegraphics[
        width=0.95\textwidth,
        trim=0 0 0 0,
        clip
    ]{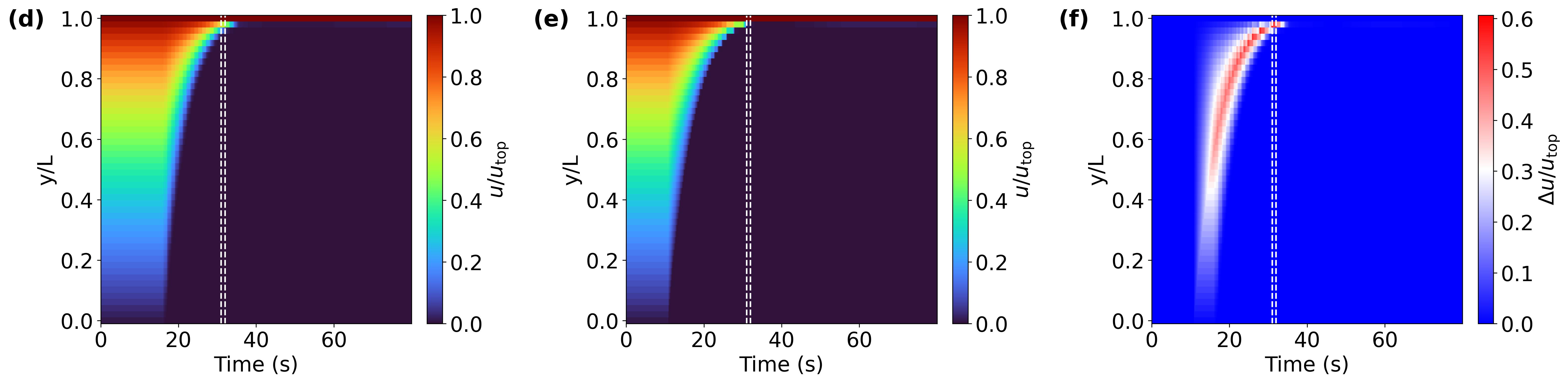}
    \caption{\textit{Top:} Ramp down/up flows of the NGF and MNGF models. The blue curve corresponds to the NGF limit without structural evolution, while the black and red curves show the down- and up-sweep MNGF flows, respectively. \textit{Bottom:} Velocity–profile comparison for a representative shear–rate sweep in the MNGF model with a dwell time of $\delta t = 1.0$~s. From left to right: (d) Down-sweep segment, (e) Up-sweep segment mirrored [in time] for visual comparison with the down-sweep flow, and (f) Flow difference between the down- and mirrored up-sweep flows, $\lvert \Delta (u/u_{\mathrm{top}}) \rvert$. Vertical dashed lines mark the time at which the highest shear stress difference in down- and up-sweep flows is observed, also marked in the flow curves of the top row. Simulation parameters are  $\rho=2450~\mathrm{kg\,m^{-3}}$, $A=0.84$, $d=8\times10^{-4}~\mathrm{m}$, $\mu_s=0.29$, $\beta=0.10$, $t_0=0.01~\mathrm{s}$, $a_{\mathrm{age}}=5.0$, $b=1.5$, $\alpha=1.0$, $\tau=20.0~\mathrm{s}$, with $\dot{\gamma}_{\min}=10^{-6}~\mathrm{s^{-1}}$, $\dot{\gamma}_{\max}=10^{3}~\mathrm{s^{-1}}$, and dwell times $\delta t=\{1,10,100\}~\mathrm{s}$.}
    \label{fig:hysteresis_velocity}
\end{figure*}

A hallmark of kinematic memory in complex fluids is rheological hysteresis~\cite{Jamali2019, Divoux2013, Radhakrishnan2017}. Experimentally measured velocity fields in thixotropic fluids and also their numerical modeling through the soft glassy rheology (SGR) model have demonstrated a strong connection between hysteretic effects and non-homogeneous flow fields~\cite{Divoux2013, Radhakrishnan2017}. Hence, we next test the MNGF in a standard ramp down/up flow protocol. Starting from a relatively large deformation rate $\dot{\gamma}_{\max}=10^{3}\,\mathrm{s^{-1}}$, the shear rate is decreased stepwise down to $\dot{\gamma}_{\min}=10^{-6}\,\mathrm{s^{-1}}$ over $N$ logarithmically spaced steps with $N=15$ points per decade and a fixed dwell time $\delta t$ per step (down-sweep). Without delay, the sweep is then reversed from $\dot{\gamma}_{\min}$ back to $\dot{\gamma}_{\max}$ using the same $(N,\delta t)$ (up-sweep). Before the cycle, samples are pre-sheared at $\dot{\gamma}_{\max}$ to erase the previous structural history and establish a reproducible initial state. In simulations, a hydrostatic pressure profile $P(y)$ is included to account for gravity, and we impose $\dot{\gamma}$ while evolving the nonlocal fluidity $g$ and structure $\lambda$. 

The results in Fig.~\ref{fig:hysteresis_velocity} clearly demonstrate that hysteresis loops as well as non-monotonic flow curves are recovered through MNGF, while NGF only recovers monotonic flow curves with no memory effects. The MNGF fluid shows strong flow hysteresis that grows as the dwell time $\delta t$ is decreased. However, non-monotonicity of flow curves appears to not significantly change as a function of dwell time. As the dwell time $\delta t$ increases, the material has more time to structurally rebuild/age at each imposed shear rate. This stronger recovery means that when the flow is resumed during the ramp up, the system resists deformation more strongly at lower shear rates. Consequently, the point of maximum stress difference between down-sweep and up-sweep, denoted $\dot{\gamma}^\star$, shifts to smaller shear rates. These findings confirm that coupling nonlocal fluidity with aging--rejuvenation dynamics captures not only steady-state rheology but also history-dependent effects inaccessible to purely viscoplastic models. As suggested by previous works of~\cite{Radhakrishnan2017, Divoux2013, Jamali2019}, non-monotonic hysteretic flow curves show strong non-homogeneities in their velocity profiles as well. To validate this, kymographs of the velocity profile for the MNGF flow with $\delta t = 1~\mathrm{s}$ [Fig.~\ref{fig:hysteresis_velocity}(b)] are presented in Fig.~\ref{fig:hysteresis_velocity}(d,e), where the velocity profile is mirrored [in time] for the up-sweep [Fig.~\ref{fig:hysteresis_velocity}(d)] flow to facilitate the visual comparison with the down sweep kymograph [Fig.~\ref{fig:hysteresis_velocity}(e)]. While the overall flow maps for the ramp down/up flows are similar, there is a clear difference that is better highlighted in the velocity differential map shown in Fig.~\ref{fig:hysteresis_velocity}(f). Note that the maximum differential between the velocity profiles in ramp down/up does not coincide with the maximum differential observed in the shear stress during down- and up-sweep flows marked by the vertical line in kymographs as well as in Fig.~\ref{fig:hysteresis_velocity}(b). This is consistent with findings of~\cite{Jamali2019}, that the timescale for macroscopic thixotropic effects measured through overall shear stress does not necessarily coincide with mesoscopic thixotropic features measured through velocity profiles. 




Another typical behavior of dense granular flows with memory effect is their delayed yielding under creep flow experiments~\cite{Coussot2002PRL, Divoux2013}, where a constant shear stress $\sigma$ is applied, and the material’s strain rate $\dot{\gamma}(t)$ is monitored as a function of time. For simple yield-stress fluids, the response is expected to be binary: flow when the applied stress exceeds the yield stress, $\sigma > \sigma_y$, and arrest for $\sigma < \sigma_y$. However, thixotropic and aging materials display richer dynamics, including slow relaxation, delayed yielding, and even transient shear banding. Each creep test here begins with a pre-shear to reset the structure, followed by two cases: one with immediate stress application (no rest) and another with a finite rest period of $t_{\text{rest}} = 20~\mathrm{s}$, allowing partial structural rebuilding. The applied stresses are varied systematically from $0.2$ to $20~\mathrm{kPa}$, covering sub-yield, near-yield, and yielded regimes. The subsequent evolution of $\dot{\gamma}(t)$ is then monitored to assess whether the material flows, arrests, or exhibits delayed yielding.

\begin{figure}[htbp]
    \centering

    \includegraphics[
        width=0.95\linewidth,
        trim=0 20 0 15,  
        clip
    ]{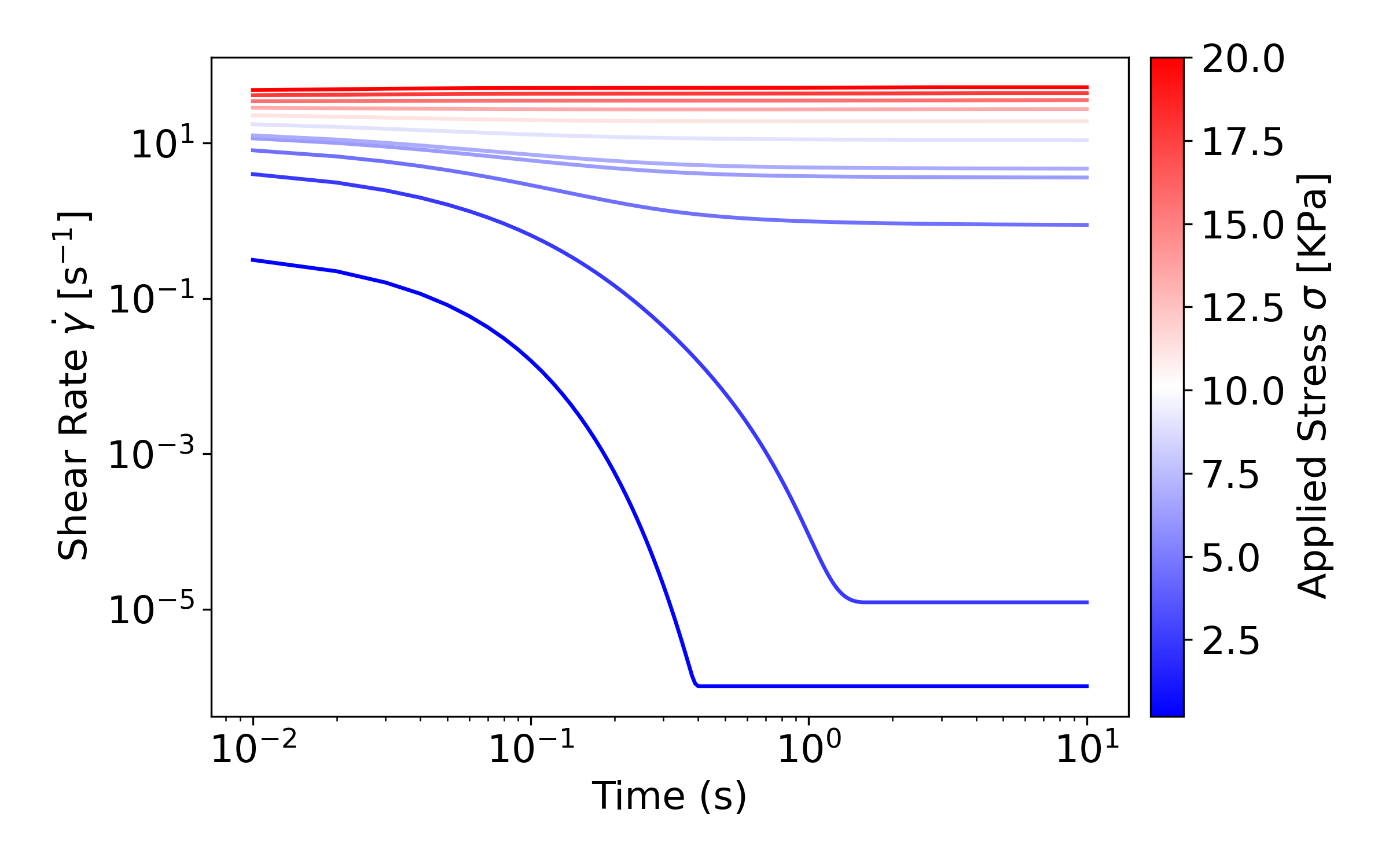}

    \vspace{0.8em}

    \includegraphics[
        width=0.95\linewidth,
        trim=0 20 0 15,
        clip
    ]{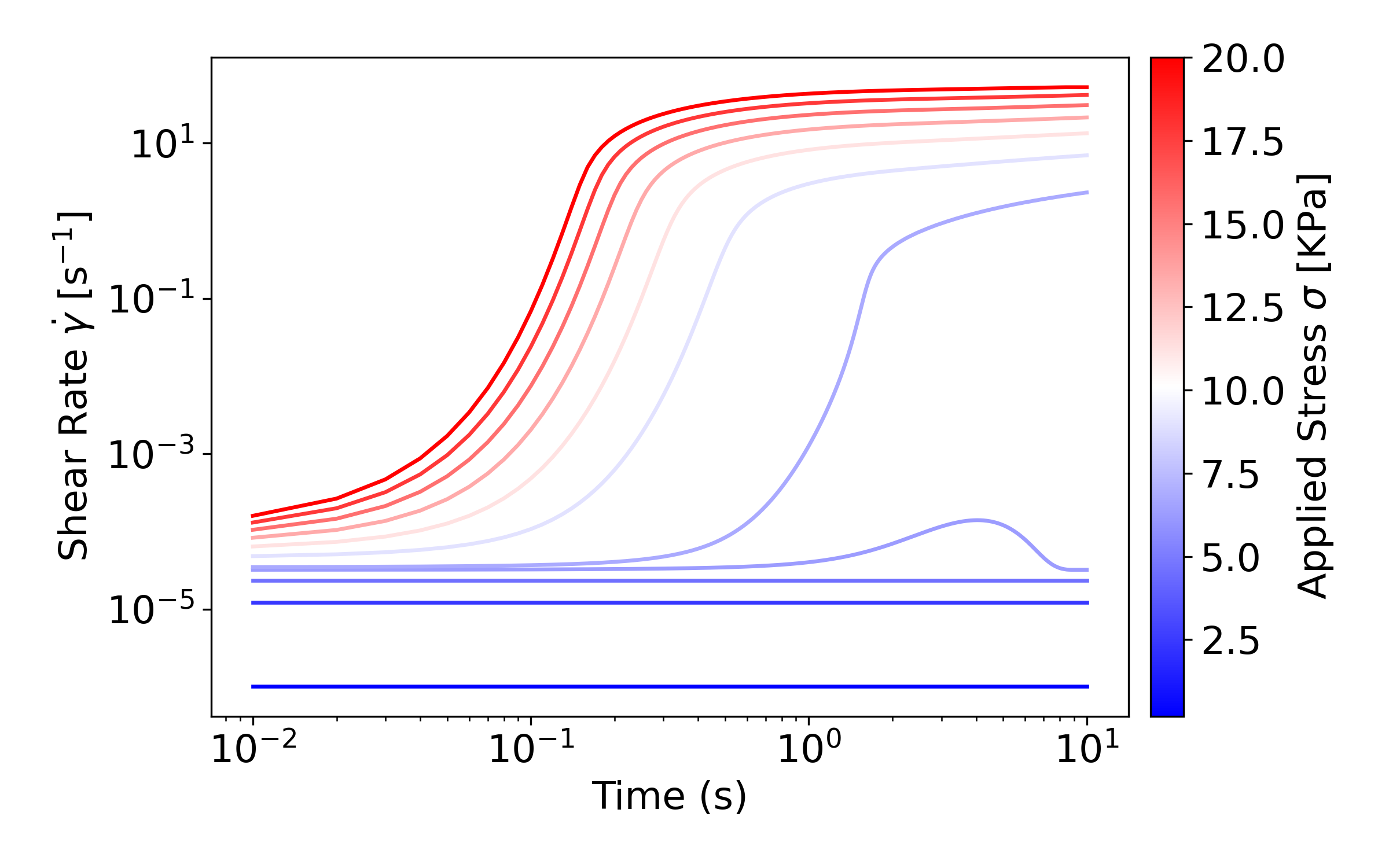}

    \caption{Creep response of the MNGF model under constant applied stress $\sigma$ (color-coded). Time evolution of the shear rate $\dot{\gamma}(t)$ (top) without rest and (bottom) after a $20~\mathrm{s}$ rest period, during which the structural parameter $\lambda$ rebuilds and increases the effective yield threshold $\sigma_y$.}
    \label{fig:creep-comparison}
\end{figure}

As evident in Fig.~\ref{fig:creep-comparison}, when no rest is introduced, $\lambda$ stays low and the system remains partially fluidized from pre-shear. For sub-yield stresses, the imposed load is insufficient to overcome the structural strength of the material. As a result, the shear rate $\dot{\gamma}(t)$ decays over time: initially, the destructuration through the pre-shear step allows some motion, but progressive structural rebuilding dominates, causing the flow to diminish. Importantly, $\dot{\gamma}$ does not vanish completely, reflecting persistent creep due to the nonlocal fluidity and incomplete arrest typical of thixotropic systems. For near-yield stresses, $\dot{\gamma}(t)$ first decreases, then passes through a shallow minimum before increasing again. This transient behavior corresponds to a competition between aging, which suppresses motion, and rejuvenation, which gradually activates flow once the applied stress is close enough to the yield threshold. The result is a delayed onset of steady flow without a sharp transition. For super-yield stresses, the applied load strongly exceeds the structural resistance. Here, $\dot{\gamma}(t)$ grows toward a steady plateau, as rejuvenation rapidly breaks down the internal structure and sustains continuous flow. In this regime, the buildup of $\lambda$ is unable to compensate, so the system fluidizes almost immediately.

When a rest period is imposed before creep, the response changes dramatically. During the rest period, the structural parameter $\lambda$ rebuilds toward unity, which represents a fully aged, solid-like state. This structural recovery strengthens the material by increasing its effective yield stress. For near-yield stresses, the creep curve develops a pronounced \textit{S}-shape. At the start of loading, the material is strongly aged (large $\lambda$), so the applied stress is insufficient to immediately fluidize it. As a result, $\dot{\gamma}(t)$ stays very small during an initial induction phase, reflecting a metastable solid-like state. Over time, localized structural breakdown reduces $\lambda$ and gradually increases fluidity $g$. Once rejuvenation overcomes aging, the material yields abruptly, and $\dot{\gamma}(t)$ accelerates sharply, producing the upward bend of the \textit{S}-curve. However, because the applied stress is only slightly above yield, rejuvenation cannot completely suppress ongoing aging. As a result, $\dot{\gamma}(t)$ does not remain at a steady plateau but gradually decreases again, indicating partial re-solidification. This interplay of induction, delayed yielding, and subsequent decay gives rise to the full \textit{S}-shaped creep response observed near the yield threshold. For sub-yield stresses, the rebuilt structure is never fully broken by the applied load, so $\dot{\gamma}(t)$ quickly decays and remains near zero throughout the test. The material stays in a solid-like arrested state, exhibiting only negligible creep due to residual fluidity. The rest-induced delayed yielding recovered here is consistent with experimental studies on concentrated colloidal pastes and bentonite suspensions, which show rest-time–dependent yielding and viscosity bifurcation~\cite{Coussot2002Viscosity}.


Here, by embedding a rate- and time-dependent structural component into the NGF framework, we introduced a memory-augmented non-local granular fluidity (MNGF) constitutive model. While the classical NGF successfully describes cooperativity, it lacks memory and therefore cannot reproduce the behaviors that are central to granular and yield-stress materials' time-dependent features. Our augmented model bridges this gap by coupling the nonlocal fluidity field $g$ with the structural parameter $\lambda$, enabling the description of time-dependent responses. Our results demonstrate that internal structural evolution plays a decisive role in describing granular fluids' kinematic memory and sensitivity to the history of deformation. This manifests in non-monotonic and hysteretic flow curves under ramp down/up experiments that involve highly non-homogeneous velocity profiles. These hysteretic features can be accurately quantified from the overall shear stress response of the fluid as well. Moreover, the proposed MNGF model reproduced a range of classical behaviors in creep experiments: persistent slow creep below yield, delayed yielding with S-shaped curves near yield, and rapid fluidization above yield. These responses are fully consistent with experimental results from dense colloidal suspensions, gels, and slurries~\cite{Coussot2002Viscosity}. Taken together, these tests strongly suggest the MNGF framework captures both nonlocal cooperativity and structural memory, unifying nonlocal granular rheology with history-dependent effects. By coupling memory to nonlocal fluidity, we show that history dependence and cooperativity cannot be treated independently. This minimal extension opens a route toward predictive modeling of time-dependent flows in amorphous materials from granular yield-stress materials to dense and jamming particulate systems and soft glassy materials.


\bibliography{apssamp}

\end{document}